\documentclass[prl,superscriptaddress,amsmath, amssymb,reprint]{revtex4-1}

\pdfoutput=1

\usepackage{graphicx}
\usepackage{epstopdf}
\usepackage[pdftex,breaklinks=true,bookmarksopen=true,bookmarksopenlevel=3,bookmarksnumbered=true,colorlinks=true,urlcolor= magenta,citecolor=blue,linkcolor=blue]{hyperref}

\begin{document}

\title{Mapping the X-Ray Emission Region in a Laser-Plasma Accelerator}

\author{S.~Corde}
\author{C.~Thaury}
\author{K.~Ta Phuoc}
\author{A.~Lifschitz}
\author{G.~Lambert}
\author{J.~Faure}
\author{O.~Lundh}
\affiliation{Laboratoire d'Optique Appliqu\'ee, ENSTA ParisTech - CNRS UMR7639 - \'Ecole Polytechnique, Chemin de la Huni\`ere, 91761 Palaiseau, France}
\author{E.~Benveniste}
\author{A.~Ben-Ismail}
\affiliation{Laboratoire Leprince Ringuet, \'Ecole Polytechnique - CNRS-IN2P3 UMR7638, Route de Saclay, 91128 Palaiseau, France}
\author{L.~Arantchuk}
\affiliation{Laboratoire de Physique des Plasmas, \'Ecole Polytechnique - CNRS UMR7648, Route de Saclay, 91128 Palaiseau, France}
\author{A.~Marciniak}
\author{A.~Stordeur}
\author{P.~Brijesh}
\author{A.~Rousse}
\affiliation{Laboratoire d'Optique Appliqu\'ee, ENSTA ParisTech - CNRS UMR7639 - \'Ecole Polytechnique, Chemin de la Huni\`ere, 91761 Palaiseau, France}
\author{A. Specka}
\affiliation{Laboratoire Leprince Ringuet, \'Ecole Polytechnique - CNRS-IN2P3 UMR7638, Route de Saclay, 91128 Palaiseau, France}
\author{V. Malka}
\affiliation{Laboratoire d'Optique Appliqu\'ee, ENSTA ParisTech - CNRS UMR7639 - \'Ecole Polytechnique, Chemin de la Huni\`ere, 91761 Palaiseau, France}

\begin{abstract}
The x-ray emission in laser-plasma accelerators can be a powerful tool to understand the physics of relativistic laser-plasma interaction. It is shown here that the mapping of betatron x-ray radiation can be obtained from the x-ray beam profile when an aperture mask is positioned just beyond the end of the emission region. The influence of the plasma density on the position and the longitudinal profile of the x-ray emission is investigated and compared to particle-in-cell simulations. The measurement of the x-ray emission position and length provides insight on the dynamics of the interaction, including the electron self-injection region, possible multiple injection, and the role of the electron beam driven wakefield.
\end{abstract}

\maketitle

Remarkable advances in relativistic laser-plasma interaction using intense femtosecond lasers have led to the development of compact electron accelerators and x-ray sources with unique properties. These sources use the very high longitudinal electric field associated with plasma waves, excited in an under-dense plasma by a relativistic laser pulse, to trap and accelerate electrons to relativistic energies~\cite{PRL1979Tajima}. For laser and plasma parameters corresponding to the bubble or blowout regimes~\cite{APB2002Pukhov, PRSTAB2007Lu}, the production of quasimonoenergetic electron beams was demonstrated~\cite{Nature2004Mangles, *Nature2004Geddes, *Nature2004Faure}. During their acceleration, these electrons  perform transverse (betatron) oscillations due to the transverse focusing force of the wakefields. This leads to the emission of bright and collimated femtosecond beams of x rays~\cite{PRL2004Rousse, PRL2004Kiselev, NatPhys2010Kneip}. Such a compact and cost effective x-ray source could contribute to the development of emerging fields such as femtosecond x-ray imaging~\cite{Science2007Gaffney}. The x-ray emission can be exploited as well to provide information on the physics of laser-plasma interaction, such as electron trajectories in the bubble~\cite{PRL2006TaPhuoc}. 

In this Letter, we demonstrate that by measuring the position and the longitudinal profile of the x-ray emission, one can determine important features of the interaction: laser pulse self-focusing, electron self-injection and possible multiple injection, as well as the role of the electron beam wakefield. The method relies on the observation, in the x-ray beam profile, of the shadow of an aperture mask adequately positioned just beyond the end of the emission region. The size of the shadow on the x-ray image permits us to determine the longitudinal position of the x-ray emission in the plasma, while the intensity gradient of the edge of the shadow yields the longitudinal profile of the emission. Because the x-ray emission position and length are closely connected to the electron injection position and the acceleration length, this measurement provides a unique insight into the interaction. Particle-in-cell (PIC) simulations are performed to analyze the experimental results.

The experiment was conducted at Laboratoire d'Optique Appliqu\'ee with the ``Salle Jaune'' Ti:Sa laser system, which delivers 0.9 Joule on target with a full width at half maximum (FWHM) pulse duration of 35 fs and a linear polarization. The laser pulse was focused inside a capillary at $3.5 \pm 1.5$ mm from the entrance, with a $f/18$ spherical mirror. The FWHM focal spot size was $22$ $\mu$m, and using the measured intensity distribution in the focal plane we found a peak intensity of $3.2\times10^{18}\:\textrm{W.cm}^{-2}$, corresponding to a normalized amplitude of $a_0=1.2$.
The target was a capillary made of two Sapphire plates with half-cylindrical grooves of diameter $d_\textrm{cap}=210$ $\mu$m and a length of 15 mm, filled with hydrogen gas at pressure ranging from 50 to 500 mbar. The target acts as a steady-state-flow gas cell~\cite{PRL2008Osterhoff}. The capillary wall surface roughness is around 1 $\mu$m, and therefore x rays cannot be reflected by the capillary wall. In addition, from our laser contrast of $10^8$, $f$ number and capillary diameter, we estimate that the pedestal intensity on the capillary wall is smaller than $10^7$ W.cm$^{-2}$, and thus no preplasma is formed before the x-ray pulse arrival.
We measured either the x-ray beam profile using an  x-ray CCD camera ($2048\times2048$ pixels, $13.5\times13.5\:\mu\textrm{m}^2$), set at $D=73.2$ cm from the capillary exit and protected from the laser light by a 20 $\mu$m Al filter, or the electron beam spectrum with a focusing-imaging spectrometer.

\begin{figure}
\includegraphics[width=8.5cm]{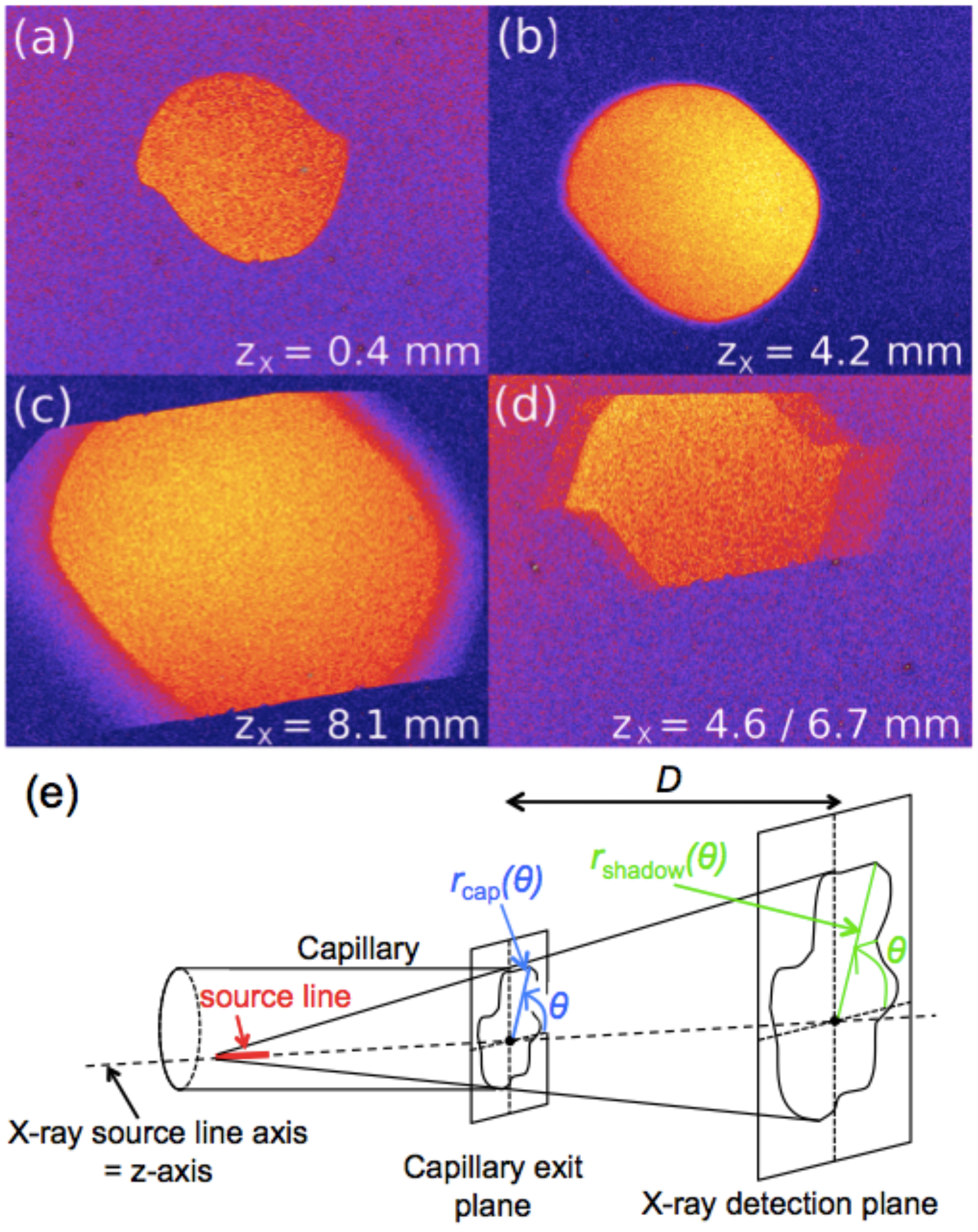}
\caption{(a)-(d) Single shot measurements of x-ray transverse profiles showing the shadow of the capillary exit, whose size permits us to deduce the emission position $z_X$ in the plasma. The asymmetric shape of the shadows is due to a slight misalignment of the two parts of the capillary. Each x-ray image has a $2.76\:\textrm{cm}\times2.13\:\textrm{cm}$ size, and the camera is situated at $D=73.2$ cm from the capillary exit. The x-ray image (d) shows an example of two separate emission positions. The horizontal edges in (c) and (d) are due to a rectangular aperture located a few centimeters after the target. (e) Schematic illustrating how the $z$ axis, and the functions $r_\textrm{cap}(\theta)$ and $r_\textrm{shadow}(\theta)$ are defined. Note that each image corresponds to different experimental conditions. In particular, (c) was obtained with a preformed plasma waveguide using a capillary discharge \cite{NatPhys2006Leemans}.}
\label{fig1}
\end{figure}

In our experiment, the betatron emission had a divergence larger than the opening angle associated with the capillary exit, which acts as the aperture mask. The x-ray beam was thus clipped by the capillary \cite{APB2011Genoud}. Figures~\ref{fig1}(a)-\ref{fig1}(c) present a sample of shadows of diverse sizes corresponding to different longitudinal positions of the source, $z_X$.

If x rays were emitted from a point source the edge of the shadow would be perfectly sharp. In contrast, for a finite source size, the edge has a finite intensity gradient which depends on the transverse and longitudinal dimensions of the source. Previous experiments reported transverse source sizes on the order of 1-2 $\mu$m~\cite{PRL2006TaPhuoc, PRE2008Albert, APL2009Mangles, NatPhys2010Kneip, OL2011Fourmaux}. In our configuration, the observed intensity gradients are much larger than those induced by such a transverse size. For $z_X = 5$ mm, a transverse source size of 1-2 $\mu$m gives the same gradient as a longitudinal extension of 100-200 $\mu$m, which gives a limit on the longitudinal resolution of the method. The gradient length is therefore dominated by the longitudinal extension of the source and x rays are considered as emitted from a line source. In the following, we use a cylindrical coordinate system $(r,\theta, z)$ whose $z$ axis is the line source axis. If $z_\textrm{entrance}=0$ corresponds to the entrance of the capillary and $z_\textrm{exit}$ to the exit, then the x-ray emission position is given for $r_\textrm{cap}(\theta) \ll r_\textrm{shadow}(\theta)$ by $z_X \simeq z_\textrm{exit}-r_\textrm{cap}(\theta)D/r_\textrm{shadow}(\theta)$, where $r_\textrm{cap}(\theta)$ [respectively $r_\textrm{shadow}(\theta)$] is the radial distance between the $z$ axis and the capillary edge (respectively the shadow edge) in the direction defined by the angle $\theta$ [see Fig. \ref{fig1}(e)], and $D$ is the distance between the capillary exit and the observation plane. For a perfectly circular capillary exit and a line source on the capillary axis, $r_\textrm{cap}(\theta)$ simplifies to $d_\textrm{cap}/2$, but a more general capillary exit shape and an arbitrary position or orientation of the line source can be represented by the function $r_\textrm{cap}(\theta)$. 

Assuming the betatron x-ray beam profile without the mask is constant on the gradient scale length (a reasonable approximation for our experimental results), the signal profile reads:
\begin{equation}
S(r, \theta)=\int_{z(r, \theta)}^{z_\textrm{exit}}\frac{dI(z^\prime)}{dz^\prime}dz^\prime,
\label{eq1}
\end{equation}
with $z(r, \theta) = z_\textrm{exit}-r_\textrm{cap}(\theta)D/r\in\left[z_\textrm{entrance},z_\textrm{exit}\right]$. In Eq. (\ref{eq1}), $dI(z^\prime)$ is the x-ray signal that originated from the emission between $z^\prime$ and $z^\prime+dz^\prime$, and $S(r, \theta)$ is the signal measured at a given position $(r, \theta)$ on the detector. 
Equation~(\ref{eq1}) can be understood as follows. For a position $(r_0, \theta_0)$ on the detector, rays coming from $z^\prime<z(r_0, \theta_0)$ are blocked at the capillary exit, and therefore the signal measured at $(r_0, \theta_0)$ is the sum of the signal emitted between $z(r_0, \theta_0)$ and $z_\textrm{exit}$.

Taking the derivative of Eq. (\ref{eq1}), the longitudinal profile of the x-ray emission $dI(z)/dz$ can be expressed as a function of the radial profile in the detection plane:
\begin{equation}
\frac{dI(z)}{dz}=-\frac{\partial S[r(z, \theta), \theta]}{\partial r}\frac{r(z, \theta)^2}{r_\textrm{cap}(\theta)D},
\label{eq2}
\end{equation}
with $r(z, \theta)=r_\textrm{cap}(\theta)D/(z_\textrm{exit}-z)$. 
If $\delta z$ and $\delta r(\theta)$ are, respectively, the characteristic emission and intensity gradient lengths, Eq.~(\ref{eq2}) leads, for $\delta z/(z_\textrm{exit}-z_X)\ll1$, to $\delta z\approx\delta r(\theta)(z_\textrm{exit}-z_X)^2/r_\textrm{cap}(\theta)D$. Thus, the measurement of the shadow size $r_\textrm{shadow}(\theta)$ and the gradient $\delta r(\theta)$ yield, respectively, the longitudinal position $z_X$ and the longitudinal length $\delta z$ of the x-ray emission. The full emission profile $dI(z)/dz$ is retrieved from $\partial S/\partial r$ using Eq. (\ref{eq2}). Further, the transverse displacement of the shadow and the asymmetry of $\delta r(\theta)$ give information on the position and orientation of the line source. For example, we observed experimentally a drift of the line source which was correlated to a low vertical drift of the laser pulse. In addition, the asymmetry in $\delta r(\theta)$, as observed in Fig. \ref{fig1}(b), cannot originate from the transverse source size and is a signature of the longitudinally induced edge intensity gradient. This confirms that the gradient length is dominated by the longitudinal extension of the source.

\begin{figure}
\includegraphics[width=8.5cm]{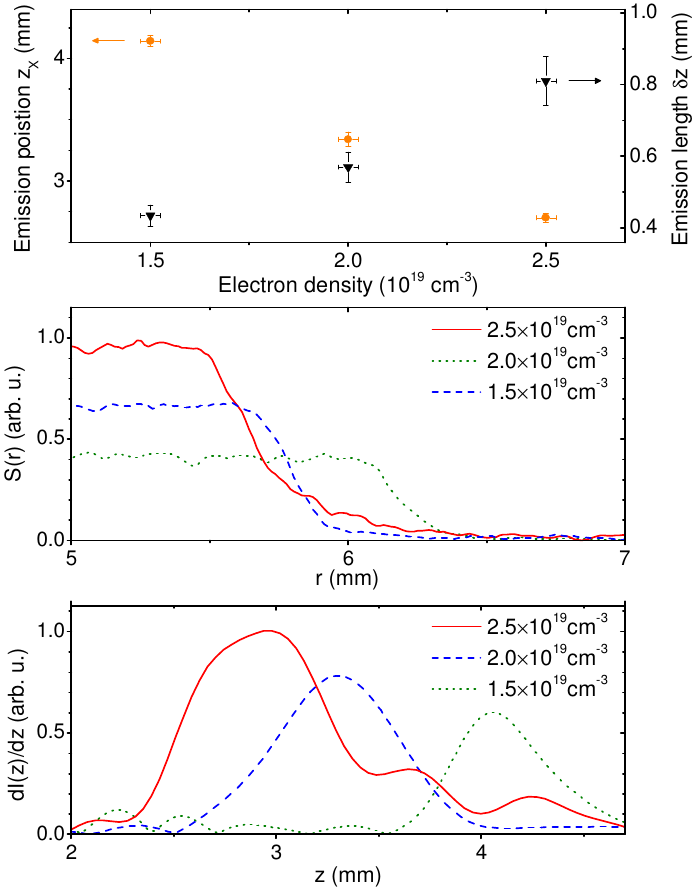}
\caption{Experimental results obtained in a steady-state-flow gas cell laser-plasma accelerator. Top: position of the beginning of the x-ray emission region $z_X$ (orange circle), and emission length $\delta z$ (defined such that 70\% of the signal is emitted in $\delta z$) (black triangle), as a function of the electron density. Each measurement point corresponds to an average over 5 to 7 shots, and the vertical error bars give the standard error of the mean. Middle: examples of raw lineouts of the edge intensity profiles $S(r)$. Bottom: the corresponding single shot x-ray emission longitudinal profiles $dI(z)/dz$, in which high frequency noise is removed by a parabolic Fourier filter [$F(k)=1-(k/k_c)^2$ for $k\leqslant k_c$, $F(k)=0$ otherwise] with a spatial cutoff frequency $k_c=16\;\text{mm}^{-1}$. The scale length associated to this filtering is $\lambda_c=390\;\mu$m.}
\label{fig2}
\end{figure}

We now apply this method to study the influence of the plasma electron density $n_e$ on the x-ray emission position $z_X$, and longitudinal extension $\delta z$. No x ray is observed for electron density $n_e< 10^{19}$~cm$^{-3}$, and the x-ray signal is increasing with $n_e$. At the density range $10^{19}$~cm$^{-3}<n_e<2.5\times 10^{19}$~cm$^{-3}$, we observed broadband electron beams with energies from 100 to 400 MeV, with sometimes some monoenergetic components, and charge in the few tens of pC range. Figure~\ref{fig2} shows the experimental results: the behaviors of $z_X$ and $\delta z$ with respect to the electron density on the top, as well as single shot raw lineouts of the edge intensity profiles $S(r)$ (middle) and the corresponding x-ray emission profiles $dI(z)/dz$ (bottom) for different densities. The error on $z_\text{exit}-z_X$, due to the uncertainty on the capillary exit diameter resulting from laser damage, is estimated to be inferior to 1\%.

\begin{figure}
\includegraphics[width=8.5cm]{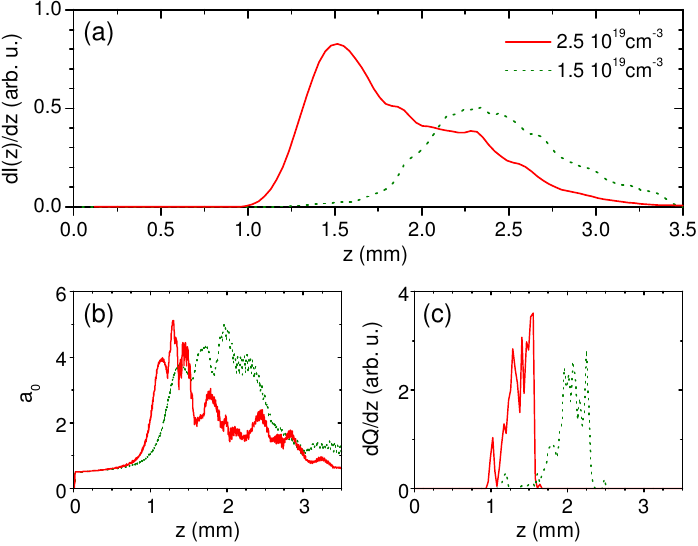}
\caption{PIC simulation results. (a) X-ray emission profiles for $n_e=2.5\times10^{19}$ cm$^{-3}$ and $n_e=1.5\times10^{19}$ cm$^{-3}$. (b) Evolution of the laser amplitude $a_0$. (c) Injected charge per unit length as a function of the position in the capillary (an electron is considered as injected when it attains 20 MeV).}
\label{fig3}
\end{figure}

To interpret these results, we performed 3D PIC simulations with the numerical code described in Ref.~\cite{calder-circ}, which uses a Fourier decomposition of the electromagnetic fields in the azimuthal direction. The first two modes are retained, which allows us to describe the linearly polarized laser field and a quasicylindrical wakefield. The normalized laser amplitude was taken as $a_0=1.1$, the FWHM focal spot width was 22 $\mu$m and the FWHM pulse duration 35 fs. The focal plane was located inside the capillary at 2.2 mm from the entrance. As the x-ray wavelengths are not resolved by the grid used in the simulations, x-ray emission was computed by using the trajectories of trapped electrons obtained from the PIC code and the Li\'enard-Wiechert fields to calculate the radiation emitted by electrons \cite{Jackson}. Figure \ref{fig3}(a) presents the calculated x-ray emission profiles for $n_e=1.5\times 10^{19}$ cm$^{-3}$ and $n_e=2.5\times 10^{19}$ cm$^{-3}$, which reproduce qualitatively the experimental profiles.

Experimentally, the position of the beginning of the x-ray emission $z_X$ varies from 4.1 mm to 2.7 mm when $n_e$ increases from $1.5\times 10^{19}$ cm$^{-3}$ to $2.5\times 10^{19}$ cm$^{-3}$. This behavior can be explained by the influence of $n_e$ on the laser propagation in the plasma. At higher density, the laser pulse self-focuses and self-steepens more quickly and towards a smaller transverse spot size~\cite{IEEE1997Esarey}, as shown by the simulated $a_0$ profiles in Fig. \ref{fig3}(b). As a result, it attains sufficiently large $a_0$ to trigger electron trapping and then x-ray emission on a smaller propagation distance. Moreover, electron self-injection is facilitated at high density, due to the stronger wakefield amplitude and the reduced wake velocity and wave-breaking threshold. Therefore, as can be seen in Fig. \ref{fig3}(b) and \ref{fig3}(c), at low density, electron trapping is delayed with respect to the beginning of the $a_0$ plateau ($a_0>4$), which contributes to an x-ray emission beginning later.

Figure~\ref{fig2} shows that the emission length $\delta z$ (in which 70\% of the signal is emitted) depends also on the electron density. It increases from 430 $\mu$m to 810 $\mu$m when the density varies from $1.5\times 10^{19}$~cm$^{-3}$ to $2.5\times 10^{19}$~cm$^{-3}$. Further, at high density, the x-ray emission length extends well beyond the dephasing and depletion lengths (the orders of magnitude are respectively $L_d\sim 200$ $\mu$m and $L_{pd}\sim500$ $\mu$m for $n_e=2.5\times10^{19}$cm$^{-3}$ considering Lu's model~\cite{PRSTAB2007Lu}).

\begin{figure}
\includegraphics[width=8.5cm]{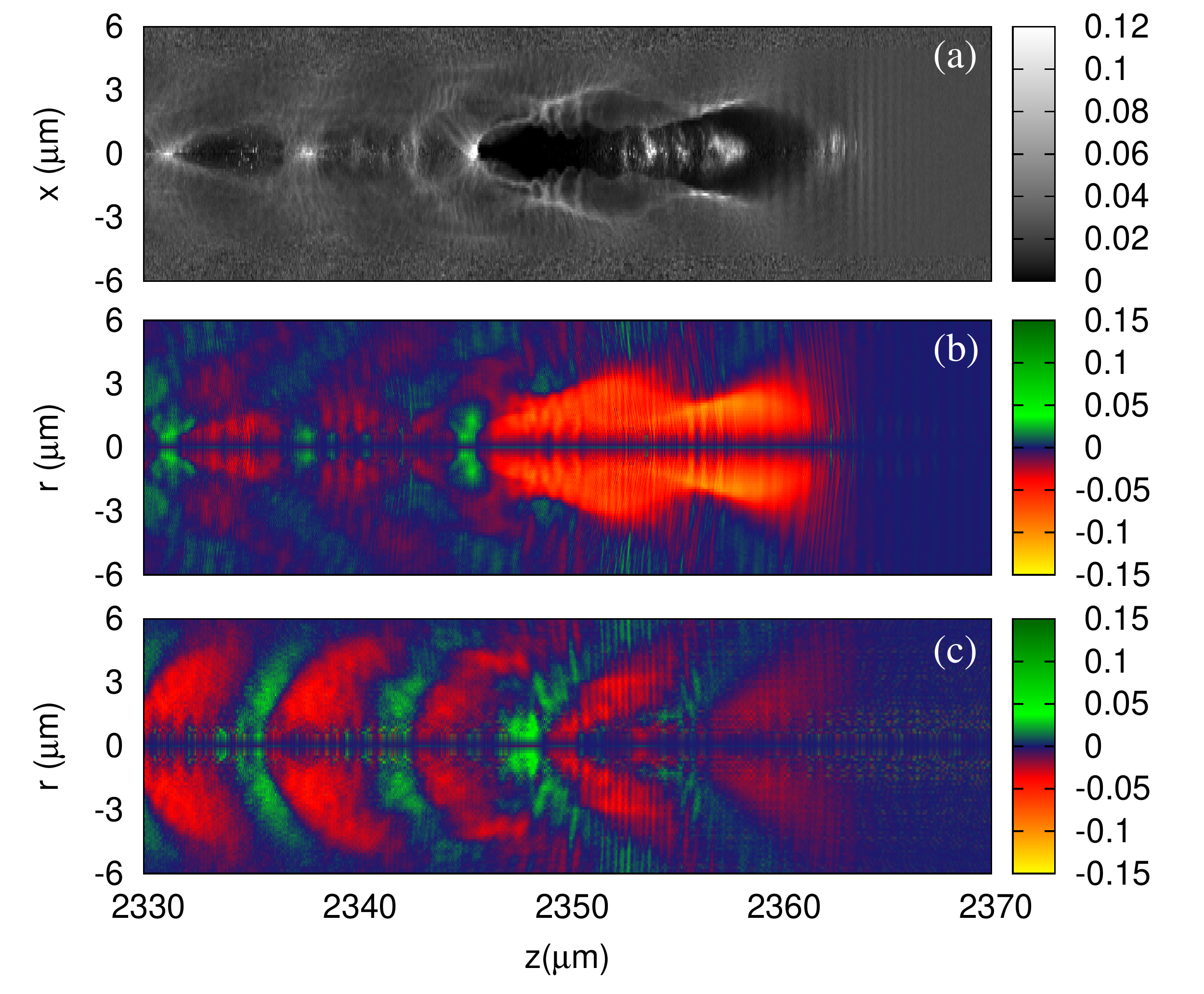}
\caption{Simulated wakefield at the end of the laser-plasma interaction for $n_e=2.5\times10^{19}$ cm$^{-3}$. (a) Electron density $n_e$, normalized  by the critical density $n_c=m_e\epsilon_0\omega^2/e^2$, where $\omega$ is the laser frequency. (b) Transverse force $F_\bot\simeq -e(E_r-cB_\theta)$, normalized by $m_ec\omega$. (c) Transverse force without the influence of the electron beam (the laser pulse is extracted and injected in a homogeneous plasma in order to calculate the wakefield induced by the laser pulse only).}
\label{fig4}
\end{figure}
To understand why x-ray emission can continue after the dephasing length  or after the fall of $a_0$, we simulated the wakefield excited by the laser pulse only (without the influence of the electron beam) at a late time. Figure \ref{fig4} shows the simulated wakefield at $z=2.35$ mm for the high density case, and the corresponding transverse force $F_\bot\simeq -e(E_r-cB_\theta)$ with and without the influence of the electron beam. At this late time, the laser pulse is unable to excite a strong transverse wakefield and the wakefield is excited mainly by the electron beam [compare Figs. \ref{fig4}(b) and Fig. \ref{fig4}(c)]. Hence, a transverse focusing wakefield is maintained by the electron beam such that electrons continue to oscillate and to emit x rays. It corresponds to a smooth transition from a laser wakefield accelerator to a plasma wakefield accelerator~\cite{PoP2010Pae} in which the wakefield is excited by a particle beam~\cite{Nature2007Blumenfeld}. Therefore, x-ray emission is no longer limited by the dephasing length or by the laser depletion. This explains why the x-ray emission length at $n_e=2.5\times10^{19}$ cm$^{-3}$ is comparable to $n_e=1.5\times10^{19}$ cm$^{-3}$ in the simulation, while the peak $a_0$ profile is significantly shorter in length in the former case. 

The behavior of $\delta z(n_e)$, particularly visible on the experimental results, could be explained by a higher normalized particle beam density $n_p/n_e$ at high density, favoring an electron beam excited transverse wakefield and a late x-ray emission.

During the experiment, multiple emission positions were observed on some shots, as shown for instance in Fig.~\ref{fig1}(d),  where x rays are emitted at $z_X=4.6$ and $z_X=6.7$ mm. This can be explained by oscillations of the laser pulse amplitude $a_0$ during its propagation in the plasma. The wakefield amplitude is sufficiently high to trap electrons only when $a_0$ is at its maximum, leading to multiple electron injection and therefore multiple emission positions.

In conclusion, we developed a novel method to map the longitudinal dependence of x-ray emission in a laser-plasma accelerator and demonstrated the possibility to measure single shot x-ray emission longitudinal profiles. This method provides detailed information on the interaction. In particular, we showed that, at high density, x-ray emission begins sooner because of the faster self-focusing and self-steepening of the laser pulse, and that the electron beam driven wakefield plays an important role in the late x-ray emission. One of the major goals for laser-plasma accelerators consists in increasing the acceleration length, either by guiding the laser pulse or using higher laser energy. In this context, this method, which can also be used with gas jets by using a small aperture near the source, will allow us to understand over which distance self-focusing and self-steepening take place, where electron injection occurs and over which distance acceleration and x-ray emission happen.

We acknowledge the Agence Nationale pour la Recherche, through the COKER project ANR-06-BLAN-0123-01, the European Research Council through the PARIS ERC project (under Contract No. 226424) and the support from EC FP7 LASERLABEUROPE/ LAPTECH Contract No. 228334 for their financial support. The authors also appreciate the contributions of J. Larour, P. Auvray and S. Hooker in the realization of the capillary unit.


\begin{thebibliography}{22}%
\makeatletter
\providecommand \@ifxundefined [1]{%
 \@ifx{#1\undefined}
}%
\providecommand \@ifnum [1]{%
 \ifnum #1\expandafter \@firstoftwo
 \else \expandafter \@secondoftwo
 \fi
}%
\providecommand \@ifx [1]{%
 \ifx #1\expandafter \@firstoftwo
 \else \expandafter \@secondoftwo
 \fi
}%
\providecommand \natexlab [1]{#1}%
\providecommand \enquote  [1]{``#1''}%
\providecommand \bibnamefont  [1]{#1}%
\providecommand \bibfnamefont [1]{#1}%
\providecommand \citenamefont [1]{#1}%
\providecommand \href@noop [0]{\@secondoftwo}%
\providecommand \href [0]{\begingroup \@sanitize@url \@href}%
\providecommand \@href[1]{\@@startlink{#1}\@@href}%
\providecommand \@@href[1]{\endgroup#1\@@endlink}%
\providecommand \@sanitize@url [0]{\catcode `\\12\catcode `\$12\catcode
  `\&12\catcode `\#12\catcode `\^12\catcode `\_12\catcode `\%12\relax}%
\providecommand \@@startlink[1]{}%
\providecommand \@@endlink[0]{}%
\providecommand \url  [0]{\begingroup\@sanitize@url \@url }%
\providecommand \@url [1]{\endgroup\@href {#1}{\urlprefix }}%
\providecommand \urlprefix  [0]{URL }%
\providecommand \Eprint [0]{\href }%
\providecommand \doibase [0]{http://dx.doi.org/}%
\providecommand \selectlanguage [0]{\@gobble}%
\providecommand \bibinfo  [0]{\@secondoftwo}%
\providecommand \bibfield  [0]{\@secondoftwo}%
\providecommand \translation [1]{[#1]}%
\providecommand \BibitemOpen [0]{}%
\providecommand \bibitemStop [0]{}%
\providecommand \bibitemNoStop [0]{.\EOS\space}%
\providecommand \EOS [0]{\spacefactor3000\relax}%
\providecommand \BibitemShut  [1]{\csname bibitem#1\endcsname}%
\let\auto@bib@innerbib\@empty
\bibitem [{\citenamefont {Tajima}\ and\ \citenamefont
  {Dawson}(1979)}]{PRL1979Tajima}%
  \BibitemOpen
  \bibfield  {author} {\bibinfo {author} {\bibfnamefont {T.}~\bibnamefont
  {Tajima}}\ and\ \bibinfo {author} {\bibfnamefont {J.~M.}\ \bibnamefont
  {Dawson}},\ }\href {\doibase 10.1103/PhysRevLett.43.267} {\bibfield
  {journal} {\bibinfo  {journal} {Phys. Rev. Lett.}\ }\textbf {\bibinfo
  {volume} {43}},\ \bibinfo {pages} {267} (\bibinfo {year} {1979})}\BibitemShut
  {NoStop}%
\bibitem [{\citenamefont {Pukhov}\ and\ \citenamefont {Meyer-ter
  Vehn}(2002)}]{APB2002Pukhov}%
  \BibitemOpen
  \bibfield  {author} {\bibinfo {author} {\bibfnamefont {A.}~\bibnamefont
  {Pukhov}}\ and\ \bibinfo {author} {\bibfnamefont {J.}~\bibnamefont {Meyer-ter
  Vehn}},\ }\href {\doibase 10.1007/s003400200795} {\bibfield  {journal}
  {\bibinfo  {journal} {Appl. Phys. B}\ }\textbf {\bibinfo
  {volume} {74}},\ \bibinfo {pages} {355} (\bibinfo {year} {2002})}\BibitemShut
  {NoStop}%
\bibitem [{\citenamefont {Lu}\ \emph {et~al.}(2007)\citenamefont {Lu},
  \citenamefont {Tzoufras}, \citenamefont {Joshi}, \citenamefont {Tsung},
  \citenamefont {Mori}, \citenamefont {Vieira}, \citenamefont {Fonseca},\ and\
  \citenamefont {Silva}}]{PRSTAB2007Lu}%
  \BibitemOpen
  \bibfield  {author} {\bibinfo {author} {\bibfnamefont {W.}~\bibnamefont
  {Lu}} \emph{et al.},\ }\href 
  {\doibase10.1103/PhysRevSTAB.10.061301} {\bibfield  {journal} {\bibinfo  {journal}
  {Phys. Rev. ST Accel. Beams}\ }\textbf {\bibinfo {volume} {10}},\
  \bibinfo {pages} {061301} (\bibinfo {year} {2007})}\BibitemShut {NoStop}%
\bibitem [{\citenamefont {Mangles}\ \emph {et~al.}(2004)\citenamefont
  {Mangles}, \citenamefont {Murphy}, \citenamefont {Najmudin}, \citenamefont
  {Thomas}, \citenamefont {Collier}, \citenamefont {Dangor}, \citenamefont
  {Divall}, \citenamefont {Foster}, \citenamefont {Gallacher}, \citenamefont
  {Hooker}, \citenamefont {Jaroszynski}, \citenamefont {Langley}, \citenamefont
  {Mori}, \citenamefont {Norreys}, \citenamefont {Tsung}, \citenamefont
  {Viskup}, \citenamefont {Walton},\ and\ \citenamefont
  {Krushelnick}}]{Nature2004Mangles}%
  \BibitemOpen
  \bibfield  {author} {\bibinfo {author} {\bibfnamefont {S.~P.~D.}\
  \bibnamefont {Mangles}} \emph{et al.},\ }\href {\doibase 10.1038/nature02939}
  {\bibfield  {journal} {\bibinfo  {journal} {Nature (London)}\ }\textbf
  {\bibinfo {volume} {431}},\ \bibinfo {pages} {535} (\bibinfo {year}
  {2004})}\BibitemShut {NoStop}%
\bibitem [{\citenamefont {Geddes}\ \emph {et~al.}(2004)\citenamefont {Geddes},
  \citenamefont {Toth}, \citenamefont {van Tilborg}, \citenamefont {Esarey},
  \citenamefont {Schroeder}, \citenamefont {Bruhwiler}, \citenamefont {Nieter},
  \citenamefont {Cary},\ and\ \citenamefont {Leemans}}]{Nature2004Geddes}%
  \BibitemOpen
  \bibfield  {author} {\bibinfo {author} {\bibfnamefont {C.~G.~R.}\
  \bibnamefont {Geddes}} \emph{et al.},\ }\href {\doibase 10.1038/nature02900} {\bibfield
  {journal} {\bibinfo  {journal} {Nature (London)}\ }\textbf {\bibinfo {volume}
  {431}},\ \bibinfo {pages} {538} (\bibinfo {year} {2004})}\BibitemShut
  {NoStop}%
\bibitem [{\citenamefont {Faure}\ \emph {et~al.}(2004)\citenamefont {Faure},
  \citenamefont {Glinec}, \citenamefont {Pukhov}, \citenamefont {Kiselev},
  \citenamefont {Gordienko}, \citenamefont {Lefebvre}, \citenamefont
  {Rousseau}, \citenamefont {Burgy},\ and\ \citenamefont
  {Malka}}]{Nature2004Faure}%
  \BibitemOpen
  \bibfield  {author} {\bibinfo {author} {\bibfnamefont {J.}~\bibnamefont
  {Faure}} \emph{et al.},\ }\href 
  {\doibase10.1038/nature02963} {\bibfield  {journal} {\bibinfo  {journal} {Nature
  (London)}\ }\textbf {\bibinfo {volume} {431}},\ \bibinfo {pages} {541}
  (\bibinfo {year} {2004})}\BibitemShut {NoStop}%
\bibitem [{\citenamefont {Rousse}\ \emph {et~al.}(2004)\citenamefont {Rousse},
  \citenamefont {Ta~Phuoc}, \citenamefont {Shah}, \citenamefont {Pukhov},
  \citenamefont {Lefebvre}, \citenamefont {Malka}, \citenamefont {Kiselev},
  \citenamefont {Burgy}, \citenamefont {Rousseau}, \citenamefont {Umstadter},\
  and\ \citenamefont {Hulin}}]{PRL2004Rousse}%
  \BibitemOpen
  \bibfield  {author} {\bibinfo {author} {\bibfnamefont {A.}~\bibnamefont
  {Rousse}} \emph{et al.},\ }\href {\doibase 10.1103/PhysRevLett.93.135005} {\bibfield
  {journal} {\bibinfo  {journal} {Phys. Rev. Lett.}\ }\textbf {\bibinfo
  {volume} {93}},\ \bibinfo {pages} {135005} (\bibinfo {year}
  {2004})}\BibitemShut {NoStop}%
\bibitem [{\citenamefont {Kiselev}\ \emph {et~al.}(2004)\citenamefont
  {Kiselev}, \citenamefont {Pukhov},\ and\ \citenamefont
  {Kostyukov}}]{PRL2004Kiselev}%
  \BibitemOpen
  \bibfield  {author} {\bibinfo {author} {\bibfnamefont {S.}~\bibnamefont
  {Kiselev}}, \bibinfo {author} {\bibfnamefont {A.}~\bibnamefont {Pukhov}}, \
  and\ \bibinfo {author} {\bibfnamefont {I.}~\bibnamefont {Kostyukov}},\ }\href
  {\doibase 10.1103/PhysRevLett.93.135004} {\bibfield  {journal} {\bibinfo
  {journal} {Phys. Rev. Lett.}\ }\textbf {\bibinfo {volume} {93}},\ \bibinfo
  {pages} {135004} (\bibinfo {year} {2004})}\BibitemShut {NoStop}%
\bibitem [{\citenamefont {Kneip}\ \emph {et~al.}(2010)\citenamefont {Kneip},
  \citenamefont {McGuffey}, \citenamefont {Martins}, \citenamefont {Martins},
  \citenamefont {Bellei}, \citenamefont {Chvykov}, \citenamefont {Dollar},
  \citenamefont {Fonseca}, \citenamefont {Huntington}, \citenamefont
  {Kalintchenko}, \citenamefont {Maksimchuk}, \citenamefont {Mangles},
  \citenamefont {Matsuoka}, \citenamefont {Nagel}, \citenamefont {Palmer},
  \citenamefont {Schreiber}, \citenamefont {Ta~Phuoc}, \citenamefont {Thomas},
  \citenamefont {Yanovsky}, \citenamefont {Silva}, \citenamefont
  {Krushelnick},\ and\ \citenamefont {Najmudin}}]{NatPhys2010Kneip}%
  \BibitemOpen
  \bibfield  {author} {\bibinfo {author} {\bibfnamefont {S.}~\bibnamefont
  {Kneip}} \emph{et al.},\ }\href 
  {\doibase10.1038/nphys1789} {\bibfield  {journal} {\bibinfo  {journal} {Nature Phys.}\
  }\textbf {\bibinfo {volume} {6}},\ \bibinfo {pages} {980} (\bibinfo {year}
  {2010})}\BibitemShut {NoStop}%
\bibitem [{\citenamefont {{Gaffney}}\ and\ \citenamefont
  {{Chapman}}(2007)}]{Science2007Gaffney}%
  \BibitemOpen
  \bibfield  {author} {\bibinfo {author} {\bibfnamefont {K.~J.}\ \bibnamefont
  {{Gaffney}}}\ and\ \bibinfo {author} {\bibfnamefont {H.~N.}\ \bibnamefont
  {{Chapman}}},\ }\href {\doibase 10.1126/science.1135923} {\bibfield
  {journal} {\bibinfo  {journal} {Science}\ }\textbf {\bibinfo {volume}
  {316}},\ \bibinfo {pages} {1444} (\bibinfo {year} {2007})}\BibitemShut
  {NoStop}%
\bibitem [{\citenamefont {Ta~Phuoc}\ \emph {et~al.}(2006)\citenamefont
  {Ta~Phuoc}, \citenamefont {Corde}, \citenamefont {Shah}, \citenamefont
  {Albert}, \citenamefont {Fitour}, \citenamefont {Rousseau}, \citenamefont
  {Burgy}, \citenamefont {Mercier},\ and\ \citenamefont
  {Rousse}}]{PRL2006TaPhuoc}%
  \BibitemOpen
  \bibfield  {author} {\bibinfo {author} {\bibfnamefont {K.}~\bibnamefont
  {Ta~Phuoc}} \emph{et al.},\ }\href
  {\doibase 10.1103/PhysRevLett.97.225002} {\bibfield  {journal} {\bibinfo
  {journal} {Phys. Rev. Lett.}\ }\textbf {\bibinfo {volume} {97}},\ \bibinfo
  {pages} {225002} (\bibinfo {year} {2006})}\BibitemShut {NoStop}%
\bibitem [{\citenamefont {Osterhoff}\ \emph {et~al.}(2008)\citenamefont
  {Osterhoff}, \citenamefont {Popp}, \citenamefont {Major}, \citenamefont
  {Marx}, \citenamefont {Rowlands-Rees}, \citenamefont {Fuchs}, \citenamefont
  {Geissler}, \citenamefont {H\"orlein}, \citenamefont {Hidding}, \citenamefont
  {Becker}, \citenamefont {Peralta}, \citenamefont {Schramm}, \citenamefont
  {Gr\"uner}, \citenamefont {Habs}, \citenamefont {Krausz}, \citenamefont
  {Hooker},\ and\ \citenamefont {Karsch}}]{PRL2008Osterhoff}%
  \BibitemOpen
  \bibfield  {author} {\bibinfo {author} {\bibfnamefont {J.}~\bibnamefont
  {Osterhoff}} \emph{et al.},\ }\href 
  {\doibase10.1103/PhysRevLett.101.085002} {\bibfield  {journal} {\bibinfo  {journal}
  {Phys. Rev. Lett.}\ }\textbf {\bibinfo {volume} {101}},\ \bibinfo {pages}
  {085002} (\bibinfo {year} {2008})}\BibitemShut {NoStop}%
\bibitem [{\citenamefont {Genoud}\ \emph {et~al.}(2011)\citenamefont {Genoud},
  \citenamefont {Cassou}, \citenamefont {Wojda}, \citenamefont {Ferrari},
  \citenamefont {Kamperidis}, \citenamefont {Burza}, \citenamefont {Persson},
  \citenamefont {Uhlig}, \citenamefont {Kneip}, \citenamefont {Mangles},
  \citenamefont {Lifschitz}, \citenamefont {Cros},\ and\ \citenamefont
  {Wahlstr\"om}}]{APB2011Genoud}%
  \BibitemOpen
  \bibfield  {author} {\bibinfo {author} {\bibfnamefont {G.}~\bibnamefont
  {Genoud}} \emph{et al.},\
  }\href {\doibase 10.1007/s00340-011-4639-4} {\bibfield  {journal} {\bibinfo
  {journal} {Appl. Phys. B}\ }\textbf {\bibinfo {volume} {105}},\
  \bibinfo {pages} {309} (\bibinfo {year} {2011})}\BibitemShut {NoStop}%
\bibitem [{\citenamefont {Albert}\ \emph {et~al.}(2008)\citenamefont {Albert},
  \citenamefont {Shah}, \citenamefont {Ta~Phuoc}, \citenamefont {Fitour},
  \citenamefont {Burgy}, \citenamefont {Rousseau}, \citenamefont {Tafzi},
  \citenamefont {Douillet}, \citenamefont {Lefrou},\ and\ \citenamefont
  {Rousse}}]{PRE2008Albert}%
  \BibitemOpen
  \bibfield  {author} {\bibinfo {author} {\bibfnamefont {F.}~\bibnamefont
  {Albert}} \emph{et al.},\ }\href {\doibase 10.1103/PhysRevE.77.056402}
  {\bibfield  {journal} {\bibinfo  {journal} {Phys. Rev. E}\ }\textbf {\bibinfo
  {volume} {77}},\ \bibinfo {pages} {056402} (\bibinfo {year}
  {2008})}\BibitemShut {NoStop}%
\bibitem [{\citenamefont {Mangles}\ \emph {et~al.}(2009)\citenamefont
  {Mangles}, \citenamefont {Genoud}, \citenamefont {Kneip}, \citenamefont
  {Burza}, \citenamefont {Cassou}, \citenamefont {Cros}, \citenamefont {Dover},
  \citenamefont {Kamperidis}, \citenamefont {Najmudin}, \citenamefont
  {Persson}, \citenamefont {Schreiber}, \citenamefont {Wojda},\ and\
  \citenamefont {Wahlstrom}}]{APL2009Mangles}%
  \BibitemOpen
  \bibfield  {author} {\bibinfo {author} {\bibfnamefont {S.~P.~D.}\
  \bibnamefont {Mangles}} \emph{et al.},\ }\href {\doibase 10.1063/1.3258022} {\bibfield
  {journal} {\bibinfo  {journal} {Appl. Phys. Lett.}\ }\textbf {\bibinfo
  {volume} {95}},\ \bibinfo {eid} {181106} (\bibinfo {year}
  {2009})}\BibitemShut {NoStop}%
\bibitem [{\citenamefont {Fourmaux}\ \emph {et~al.}(2011)\citenamefont
  {Fourmaux}, \citenamefont {Corde}, \citenamefont {Ta~Phuoc}, \citenamefont
  {Lassonde}, \citenamefont {Lebrun}, \citenamefont {Payeur}, \citenamefont
  {Martin}, \citenamefont {Sebban}, \citenamefont {Malka}, \citenamefont
  {Rousse},\ and\ \citenamefont {Kieffer}}]{OL2011Fourmaux}%
  \BibitemOpen
  \bibfield  {author} {\bibinfo {author} {\bibfnamefont {S.}~\bibnamefont
  {Fourmaux}} \emph{et al.},\ }\href {\doibase 10.1364/OL.36.002426} {\bibfield
  {journal} {\bibinfo  {journal} {Opt. Lett.}\ }\textbf {\bibinfo {volume}
  {36}},\ \bibinfo {pages} {2426} (\bibinfo {year} {2011})}\BibitemShut
  {NoStop}%
\bibitem [{\citenamefont {{Lifschitz}}\ \emph {et~al.}(2009)\citenamefont
  {{Lifschitz}}, \citenamefont {{Davoine}}, \citenamefont {{Lefebvre}},
  \citenamefont {{Faure}}, \citenamefont {{Rechatin}},\ and\ \citenamefont
  {{Malka}}}]{calder-circ}%
  \BibitemOpen
  \bibfield  {author} {\bibinfo {author} {\bibfnamefont {A.~F.}\ \bibnamefont
  {{Lifschitz}}} \emph{et al.},\ }\href
  {\doibase 10.1016/j.jcp.2008.11.017} {\bibfield  {journal} {\bibinfo
  {journal} {J. Comput. Phys.}\ }\textbf {\bibinfo {volume} {228}},\ \bibinfo
  {pages} {1803} (\bibinfo {year} {2009})}\BibitemShut {NoStop}%
\bibitem [{\citenamefont {Jackson}(2001)}]{Jackson}%
  \BibitemOpen
  \bibfield  {author} {\bibinfo {author} {\bibfnamefont {J.~D.}\ \bibnamefont
  {Jackson}},\ }\href@noop {} {\emph {\bibinfo {title} {Classical
  Electrodynamics}}},\ \bibinfo {edition} {3rd}\ ed.\ (\bibinfo  {publisher}
  {Wiley},\ \bibinfo {address} {New York},\ \bibinfo {year} {2001})\BibitemShut
  {NoStop}%
\bibitem [{\citenamefont {{Esarey}}\ \emph {et~al.}(1997)\citenamefont
  {{Esarey}}, \citenamefont {{Sprangle}}, \citenamefont {{Krall}},\ and\
  \citenamefont {{Ting}}}]{IEEE1997Esarey}%
  \BibitemOpen
  \bibfield  {author} {\bibinfo {author} {\bibfnamefont {E.}~\bibnamefont
  {{Esarey}}} \emph{et al.},\ }\href
  {\doibase 10.1109/3.641305} {\bibfield  {journal} {\bibinfo  {journal} {IEEE
  J. Quantum Electron.}\ }\textbf {\bibinfo {volume} {33}},\ \bibinfo {pages}
  {1879} (\bibinfo {year} {1997})}\BibitemShut {NoStop}%
\bibitem [{\citenamefont {Pae}\ \emph {et~al.}(2010)\citenamefont {Pae},
  \citenamefont {Choi},\ and\ \citenamefont {Lee}}]{PoP2010Pae}%
  \BibitemOpen
  \bibfield  {author} {\bibinfo {author} {\bibfnamefont {K.~H.}\ \bibnamefont
  {Pae}}, \bibinfo {author} {\bibfnamefont {I.~W.}\ \bibnamefont {Choi}}, \
  and\ \bibinfo {author} {\bibfnamefont {J.}~\bibnamefont {Lee}},\ }\href
  {\doibase 10.1063/1.3522757} {\bibfield  {journal} {\bibinfo  {journal}
  {Phys. Plasmas}\ }\textbf {\bibinfo {volume} {17}},\ \bibinfo {eid} {123104}
  (\bibinfo {year} {2010})}\BibitemShut {NoStop}%
\bibitem [{\citenamefont {Blumenfeld}\ \emph {et~al.}(2007)\citenamefont
  {Blumenfeld}, \citenamefont {Clayton}, \citenamefont {Decker}, \citenamefont
  {Hogan}, \citenamefont {Huang}, \citenamefont {Ischebeck}, \citenamefont
  {Iverson}, \citenamefont {Joshi}, \citenamefont {Katsouleas}, \citenamefont
  {Kirby}, \citenamefont {Lu}, \citenamefont {Marsh}, \citenamefont {Mori},
  \citenamefont {Muggli}, \citenamefont {Oz}, \citenamefont {Siemann},
  \citenamefont {Walz},\ and\ \citenamefont {Zhou}}]{Nature2007Blumenfeld}%
  \BibitemOpen
  \bibfield  {author} {\bibinfo {author} {\bibfnamefont {I.}~\bibnamefont
  {Blumenfeld}} \emph{et al.},\ }\href 
  {\doibase10.1038/nature05538} {\bibfield  {journal} {\bibinfo  {journal} {Nature
  (London)}\ }\textbf {\bibinfo {volume} {445}},\ \bibinfo {pages} {741}
  (\bibinfo {year} {2007})}\BibitemShut {NoStop}%
\bibitem [{\citenamefont {Leemans}\ \emph {et~al.}(2006)\citenamefont
  {Leemans}, \citenamefont {Nagler}, \citenamefont {Gonsalves}, \citenamefont
  {Toth}, \citenamefont {Nakamura}, \citenamefont {Geddes}, \citenamefont
  {Esarey}, \citenamefont {Schroeder},\ and\ \citenamefont
  {Hooker}}]{NatPhys2006Leemans}%
  \BibitemOpen
  \bibfield  {author} {\bibinfo {author} {\bibfnamefont {W.~P.}\ \bibnamefont
  {Leemans}} \emph{et al.},\ }\href {\doibase 10.1038/nphys418}
  {\bibfield  {journal} {\bibinfo  {journal} {Nature Phys.}\ }\textbf {\bibinfo
  {volume} {2}},\ \bibinfo {pages} {696} (\bibinfo {year} {2006})}\BibitemShut
  {NoStop}%
\end{thebibliography}
\end{document}